\documentclass[12pt]{article}
\usepackage{geometry}
\usepackage{color}
\usepackage{xcolor}
\usepackage{a4}
\usepackage{graphicx}
\usepackage{epsf}
\usepackage{amsmath}
\usepackage{amssymb}
\usepackage{cite}
\usepackage{color}
\usepackage{hyperref}
\usepackage{bibentry}
\newcommand{\be}{\begin{equation}}
\newcommand{\ee}{\end{equation}}

\newcommand{\Rmnum}[1]{\expandafter\@slowromancap\romannumeral #1@}
\newcommand{\bea}{\begin{eqnarray}}
\newcommand{\eea}{\end{eqnarray}}
\begin{document}
\begin{titlepage}
\title{Self-gravitating fluid systems and galactic dark matter} 
\author{} 
\date{
Uddipan Banik, Dipanjan Dey, Kaushik Bhattacharya, \\Tapobrata Sarkar
\thanks{\noindent 
E-mail:~ uddipan, deydip, kaushikb, tapo@iitk.ac.in} 
\vskip0.4cm 
{\sl Department of Physics, \\ 
Indian Institute of Technology,\\ 
Kanpur 208016, \\ 
India}} 
\maketitle 
\abstract{We study gravitational collapse with anisotropic pressures, whose end stage can mimic space-times that are seeded by galactic
dark matter. To this end, we identify a class of space-times (with conical defects) that can arise out of such a collapse process, and 
admit stable circular orbits at all radial distances. These have a naked singularity at the origin. 
An example of such a space-time is seen to be the Bertrand space-time discovered by 
Perlick, that admits closed, stable orbits at all radii. Using relativistic two-fluid models, we show that our galactic space-times might indicate 
exotic matter, i.e one of the component fluids may have negative pressure for a certain asymptotic fall off of the associated mass density,
in the Newtonian limit. We complement this analysis by studying some simple examples of Newtonian two-fluid systems, and compare
this with the Newtonian limit of the relativistic systems considered.}  
\end{titlepage}
\bigskip

\section{Introduction and motivation}

Understanding the nature of galactic dark matter has been at the focus of research over the last several decades, and possibly pose one of the
biggest challenges for the future. Several empirical proposals have been put forward regarding its properties, largely based on experimental data on 
galactic rotation curves, gravitational lensing, etc. which have been tested with varying degrees of success. In this context, 
it is natural to ask if Einstein's General theory of Relativity (GR) \cite{Weinberg}, \cite{Wald} should play a role in such an 
understanding, and over the years a substantial amount of literature has been devoted to the study of possible relativistic effects of dark matter. 

If GR studies of galactic dark matter are to be taken seriously, then one should be able to obtain possible dark matter candidate (non-vacuum) 
space-times via a gravitational collapse process. 
The question then is, what should be the nature of the metric of such space-times. In the GR literature, this has been 
addressed previously, see e.g \cite{Roberts}. A popular method in the GR literature is to assume a flat rotation curve, and derive the metric
by setting the circular velocity (obtained via GR) to be a constant (see also \cite{Sayan}). As we will mention in sequel, the inherent problem 
in defining a circular velocity in GR is that it is observer dependent, and in principle one has to think of a series of stationary Lorentzian observers
at the location of the celestial objects. 

In  \cite{BST1},\cite{BST2}, (see also \cite{BST3}) we had taken a slightly different approach to the problem. Using the empirical fact that (at least away from the
galactic centre) celestial bodies often follow circular trajectories, we had proposed that galactic space-times should have the property that these support 
closed stable trajectories at all radial distances. This was indeed a phenomenological model. Such space-times are however known in the literature, and
were discovered by Perlick \cite{Perlick}, who called them Bertrand space-times (BSTs), as these are analogues of space-time that satisfy Bertrand's
famous theorem of classical mechanics. Using a phenomenological definition of the circular velocity in lines with \cite{Roberts}, we were 
able to show that such models gave excellent fits to experimental data on galactic rotation curves for low surface brightness galaxies, 
and gravitational lensing. In addition, in an
appropriate non-relativistic limit, it was shown that the energy density of BSTs match with the Navarro-Frenk-White \cite{NFW} or the
Hernquist \cite{Hernquist} density profiles that are popular in the dark matter literature. 

A salient feature of BSTs is that they have a central singularity that is naked, i.e not covered by an event horizon. In this context, we recall that
in the absence of a concrete proof of the cosmic censorship conjecture, such a singularity cannot be a-priori ruled out. It is also possible that the 
nature of such singularities might be modified due to possible quantum effects close to the centre. We also recall that both the NFW and the
Hernquist profiles (for that matter most known profiles of galactic dark matter halos) have singular behaviour near the centre, which is avoided by
a cut-off introduced by hand. 

The purpose of the present work is to build on our previous work of \cite{BST1},\cite{BST2}, \cite{BST3}. In particular, we address the following issues that
we believe are of significant interest in the study of galactic dark matter halos : 
\begin{itemize}
\item
What are possible space-times that are formed out of gravitational collapse with a central naked singularity, that supports stable circular orbits at all radii. 
In particular, these space-times should have pressure anisotropy, and to relate to realistic galactic systems, we demand that 
these should be matched to an external Schwarzschild solution. 
\item
For such (non-vacuum) space-times, what is the nature (i.e equations of state) of the anisotropic fluid(s) that source them. 
\item
What are the Newtonian (i.e non-relativistic) limits of such fluids. 
\item
Can such Newtonian limits be compared with purely Newtonian fluids that satisfy Navier-Stokes, Poisson and continuity equations. 
\end{itemize}
Let us motivate these issues. As mentioned, at least away from the galactic centre, celestial objects move in roughly circular trajectories. It is
therefore a reasonable assumption that space-times seeded by galactic dark matter should support stable circular orbits at all radii. Whether these are 
closed or not under perturbation is a more involved question to answer, and has been addressed in \cite{Perlick}. As we will see later,
our analysis uncovers a class of possible space-times, out of which BSTs are a special case. 
In a GR setup, one would expect that such space-times should be the end stage of a gravitational collapse process. 
Once such space-times are identified, one would ideally like to know the nature of the fluid matter that sources them. In this context, the second point
above finds significance. Finally, in order to connect to experimental data, one has to understand the Newtonian limits of such models, and also
compare these with purely Newtonian fluids, which are motivations for the last two points listed above. 

In the first part of this paper (section 2.1), we set up the basic framework of gravitational collapse to a naked singularity in the 
context of GR. This is done by generalising the recent work of \cite{JMN1},\cite{JMN2} to include anisotropy in the component fluid pressures.
Section 2.2 deals with specific examples of space-times (with a central naked singularity) that admit stable circular orbits at all radii. In section 3, 
we model a class of these space-times by anisotropic fluids. Section 4 is devoted to the study of the non-relativistic limits
of these fluid models. Finally, in section 5, we phenomenologically consider purely Newtonian two-fluid models to reconcile with 
the results of sections 3 and 4.

Our main findings in this paper are the following. Firstly, we find that the Joshi-Malafarina-Narayan (JMN) space-time \cite{JMN1} 
and the BST are special cases of a generic
class of space-times that can be formed due to gravitational collapse, which have a conical defect, and which support stable circular orbits at 
all radii. Secondly, we show that both these anisotropic space-times can be solved using a two-fluid relativistic model, and for the BST, it is seen
that one of the component fluids is necessarily exotic, i.e has negative pressure akin to dark energy. Thirdly, we find that this latter feature is 
possibly due to the nature of the fall-off of the mass density at asymptotic infinity, in a Newtonian limit. This last feature is seen to be similar to purely
Newtonian composite fluids with some simple forms of the velocity of the component fluids, with a few differences.

To make contact with existing literature, we note here that over the last few decades, several models of dark matter have been proposed, an important 
one being the hypothesis that dark matter
is a relativistic fluid with pressure, with a definite equation of state. In one of the first works in this direction, the authors of \cite{Sayan} 
considered a scenario in which galaxy halos were modelled as dark matter with anisotropic pressure and used gravitational lensing data to
determine the dark matter equation of state. In \cite{Serra}, a similar hypothesis was tested by observations from some
galaxy clusters like the Coma cluster and CL0024 and the authors attempted to obtain bounds on the dark matter equation of state parameter,
defined as the ratio of the sum of the pressures to the mass density. Polytropic equations of states in such situations have also been commonly 
studied in the literature, see e.g \cite{Saxton}. 

Throughout this paper, we work in units where we set the speed of light and the Newton's constant to be unity. We will begin our analysis by 
addressing the issue of gravitational collapse in GR with anisotropy. 

\section{General Relativistic collapse with anisotropy}

Gravitational collapse of spherically symmetric, pressure-less matter (i.e dust) was studied by Oppenheimer and Snyder in the framework of general 
relativity in \cite{OS}. Ever since, the subject continued to receive a lot of attention. In \cite{MS}, Misner and Sharp studied the problem analytically in a more 
general situation, with a non-vanishing pressure gradient. These equations were numerically solved by May and White \cite{MW}. More recently, 
Joshi et. al \cite{JMN1}, \cite{JMN2} have approached the problem from a slightly different perspective and have shown that the end state of collapse
of matter that admit pressure gradients might be black holes, or naked singularities. As mentioned, in the absence of
a concrete proof of the cosmic censorship conjecture that might render such singularities as unphysical, naked singularities remain an important 
arena of investigation of aspects of general relativity, although there are debates in the literature about whether such singularities arise due to specific
symmetries of models which might be violated in more realistic situations. It is also worth mentioning that these singularities are classical, in the sense that
quantum corrections might modify their nature, although it is difficult to speculate on the exact nature of such corrections. 

\subsection{Anisotropic collapse to a naked singularity : basics}

We consider naked singularities, that are spherically symmetric space times containing a central singularity
which does not have an event horizon. Our purpose in this section would be to review and analyze 
some aspects of gravitational collapse that lead to naked singularities. In particular, we extend the analysis of \cite{JMN1} to include situations where the
underlying fluid is {\it anisotropic}. This allows us to make general comments regarding the nature of the radial and tangential pressures in the same, which
will be useful for us in the next section. 

We will start with a spherically symmetric metric describing gravitational collapse that can be written in diagonal form as 
$ds^2 = -e^{2\nu}dt^2 + e^{2\lambda}dr^2 + R^2d\Omega^2$, where $\nu$, $\lambda$ and $R$ are functions of the (co-moving) radial coordinate $r$ and
time $t$, with $d\Omega^2$ being the standard metric on the unit two-sphere. It is more convenient to write the metric as 
\begin{equation}
ds^2 = -e^{2\nu}dt^2+\frac{R^{\prime 2}}{G}dr^2 + R^2 d\Omega^2~,
\label{metric}
\end{equation}
where $G=G(r,t)$. Einstein's equations then implies that the energy density, radial pressure are respectively
\begin{equation}
\rho = \frac{F^{\prime}}{R^2R^\prime}~~,~~ P_{r}= -\frac{\dot{F}}{R^{2}\dot{R}}~
\label{EE1}
\end{equation}
where a dot represents a time derivative and the prime represents a derivative with respect to the radial variable $r$, and 
\begin{eqnarray}
F=R\left(1-G+e^{-2\nu}\dot{R}^2\right)~,
\label{EE2_f}
\end{eqnarray}
which specifies the amount of matter enclosed by a shell located at $r$, and is called the Misner-Sharp mass. We will be interested in the case of
asymmetric fluids, in which case the expression for the tangential pressure can be obtained from the expression for the conservation of the 
stress tensor. From the Einstein equations we also obtain 
\begin{equation}
\dot{G} = 2\frac{\nu^{\prime}}{R^{\prime}}\dot{R}G~.
\label{EE2}
\end{equation} 
We thus have a total of $7$ variables and $5$ equations, and hence have the freedom to specify any two functions. 

From eq.(\ref{EE2_f}), one can define an effective potential ($V_{eff}$):
\begin{equation}
V_{eff}= -\dot{R}^2=-e^{2 \nu(r,R)}\left(\frac{F(r,R)}{R}+G(r,R)-1\right)~,
\end{equation}
so that, in order to achieve an equilibrium condition, we need $\dot{R}=0$, $\ddot{R}=0$. Note that when $\dot{R}=0, \ddot{R}>0$ we have a bouncing condition,
and $\dot{R}=0, \ddot{R}<0$ describes a collapse.
Let us first consider the case of collapsing dust, for which $P_{r}=P_{\theta}=P_{\phi}=0$. Then we get from eqs.(\ref{EE1}), eq.(\ref{EE2_f}) and (\ref{EE2}), 
$\dot{F}=0, \nu^{\prime}=0$, $\dot{G}=0$. So $G$ and $F$ become time-independent, i.e $R$ independent, from our previous discussion. Hence, we now have
$\dot{R}^2=\left(\frac{F(r)}{R}+G(r)-1\right)$ for pressure less dust. So, for this case, after collapse starts ($\dot{R}<0$), 
$\dot{R}$ will always be negative as it is independent of time (since $\ddot{R} = -F(r)/(2\dot{R}^2)$). 
Hence for the dust like solution we always obtain a black-hole as a final state of collapsing metric.

Hence, we need presence of finite pressure in the fluid to balance the gravitational pull, and if the system can equilibriate ($\ddot{R}=0$),
we will have $\dot{R}=\ddot{R}=0$, which translates to $V_{eff}=V_{eff,R}=0$. 
The derivative of the effective potential with respect to $R$ is given by
\begin{equation}
V_{eff,R}=-2\nu_{,R}e^{2\nu}\left(\frac{F}{R}+G-1\right)+e^{2\nu}\left(\frac{F}{R^2}-\frac{F_{,R}}{R}-G_{,R}\right)\,.
\label{equil}
\end{equation}
Here $X_{,R}$ specifies a derivative of the quantity $X(r,R)$ with respect to the variable $R$. We  will focus on the case where equilibrium is achieved in infinite time, i.e 
\begin{eqnarray}
&~& R(r,t)\xrightarrow{t\rightarrow\infty}R_{e}(r),~ F(r,R)\xrightarrow{t\rightarrow\infty}F_{e}(r)\equiv F(r,R_{e}(r))\nonumber\\
&~& \nu(r,R)\xrightarrow{t\rightarrow\infty}\nu_{e}(r)\equiv \nu(r,R_{e}(r)),~G(r,R)\xrightarrow{t\rightarrow\infty}G_{e}(r)\equiv G(r,R_{e}(r))
\end{eqnarray}
At equilibrium as $V_{eff}(r,R_{e})=0$ and $V_{eff,R}(r,R_{e})=0$ we get ,
\begin{equation}
G_{e}(r)=1-\frac{F_{e}}{R_{e}},~(G_{,R})_{e}=\frac{F_{e}}{R_{e}^2}-\frac{(F_{,R})_{e}}{R_{e}}
\end{equation}
To simplify the notation, let us denote $R_{e}(r)=q$, so $F_{e}(r)=F(q), \nu_{e}(r)=\phi(q), G_{e}(r)=1-\frac{F}{q}$ in equilibrium.

From the second of eq.(\ref{EE1}),  we can write $P_{r}$ as (remembering that ${\dot F}=F_{,R}{\dot R}$ and using eq.(\ref{equil}))
\begin{equation}
P_{r}=\frac{2\phi_{,q}}{\rho}G(q)-\frac{F(q)}{q^{3}}=\frac{2\phi_{,q}}{q}\left(1-\frac{F(q)}{q}\right)-\frac{F(q)}{q^{3}}~.
\label{pr}
\end{equation}
In the above equation $\phi_{,q}$ represents a derivative of $\phi$ with respect to $q$. For anisotropic fluids, the 
Tolman-Oppenheimer-Volkoff equation can be written as \cite{Bowers} :
\begin{equation}
P_{r,q}=-(\rho+P_{r})\phi_{,q}+\frac{2}{q}(P_{\theta}-P_{r})~,
\label{tov}
\end{equation}
hence eqs.(\ref{pr}) and (\ref{tov}) can be used to compute $P_{\theta}$. 

It can then be verified easily that the collapsing metric of eq.(\ref{metric}) will tend to the following after a long time ($d\Omega^2 = 
q^2(d\theta^2 + \sin^2\theta d\phi^2)$):
\begin{equation}
ds^2 = -e^{2\phi(q)}dt^2+\frac{dq^2}{G(q)}+q^2d\Omega^2~,
\label{finalmetric}
\end{equation}
where it has to be remembered that $q\leq q_{b}\equiv R_{e}(r_{b})$, with $r_{b}$ being the matching radius with an external Schwarzschild metric. 
The radial and tangential pressures obtained from eq.(\ref{finalmetric}) can be shown to match with the ones obtained via the collapse scenario,
via equations (\ref{pr}) and (\ref{tov}). 

In order to obtain physically sensible space times, we will now need to match the metric of eq.(\ref{finalmetric}) with an external Schwarzschild solution. 
During collapse, before the fluid equilibrates itself, if the matching radius($r_{b}$) becomes lesser than Schwarzschild radius($r_{s}$) then the final result will 
be a black-hole. When $r_{b}$ is always greater than $r_{s}$ after a long time, then the system tends to equilibrate itself without black hole formation. 
This is the case we will be interested in here. At the end of the collapse process, there can be a singularity at the centre, which is essentially a very high 
energy density region. Whether it is naked or covered by event horizon depends on whether $r_{b}$ is greater or lesser than $r_{s}$ respectively.
Clearly, to match the metric of eq.(\ref{finalmetric}) to a Schwarzschild solution (of mass $M$), we require 
\begin{equation}
1-\frac{2M}{q_{b}}=G(q_{b})=1-\frac{F(q_{b})}{q_{b}}~.
\end{equation}
Having elucidated the basic setup of gravitational collapse to a naked singularity, we will now require to specify the forms of $F(q)$ and $\phi(q)$ 
for concrete examples of such spacetimes. This is what we will study in the next subsection. 

\subsection{Anisotropic collapse to naked singularities : examples}

In this subsection, we will study examples of a class of spacetimes that have naked singularities at the centre, and can arise out of a collapse 
process described in subsection (2.1). In order to connect to possible galactic spacetimes, we make some simplifying assumptions. 

We will focus on the case $\frac{F(q)}{q}={\rm constant}\equiv(1-\beta^2)$ for $q\leq q_{b}$. These are space-times with a conical defect and
constitute the simplest model for the end stage of gravitational collapse. In this case, 
at the junction we have $F(q_{b})=2M=(1-\beta^2)q_{b}$. 
To maintain the signature of the Schwarzschild solution, we require $\frac{2M}{q_{b}}<1 $.  Hence, we obtain 
$0<\beta^2<1$ with $G(q)=\beta^2$ specifying the conical defect. 

In order to specify the form of the function $\phi(q)$, we will demand that each point admits a stable circular orbit. 
The motivation for this is that we would finally like to focus on space-times that can be bonafide candidates for galactic dark matter, 
and for such space-times, this assumption is reasonable.  Whether these orbits are closed under small perturbations is a more 
complicated issue. Such space-times have
been studied in the literature by Perlick \cite{Perlick} and are called Bertrand space-times (BSTs), since these are general relativistic generalisations
of Bertrand's theorem of classical mechanics. In \cite{Perlick}, it was shown that a static, spherically symmetric space-time is a BST if
there exists a circular trajectory through each point, and if an initial condition that is sufficiently close to the circular orbit gives a periodic
orbit. We will mainly focus on the first condition, i.e demand that there exist stable circular trajectories at each point in our space-time. 

The two conditions above, along with the weak energy condition and the fact that our solution will be matched to an external 
Schwarzschild one will be used to put some constraints on the form of $\phi(q)$, as we now show. 

Without loss of generality, we choose to work on the equatorial plane ($\theta = \pi/2$), then for a metric with a general form 
$ds^2 = g_{tt}(q)dt^2 + g_{qq}(q)dr^2 + q^2d\Omega^2$, it can be shown that the timelike equatorial geodesics satisfy 
\begin{equation}
{\dot q}^2 + V(q) = 0,~~~V(q) = \frac{1}{g_{qq}(q)}\left[\frac{E^2}{g_{tt}(q)} + \frac{L^2}{q^2} + 1\right],
\label{genmotion}
\end{equation}
where $E$ and $L$ are the conserved energy and angular momentum respectively, per unit mass, that arise due to the 
fact that $\partial_t$ and $\partial_{\phi}$ are Killing vectors, so that $E=g_{tt}{\dot t}$ and $L=g_{\phi\phi}{\dot \phi}$ are conserved quantities. For circular
trajectories that are stable, we should impose $V(q) = V'(q) = 0$. These equations can then be used to determine $E$ and $L$, which
can be used to verify the positivity of $V''(q)$, needed for stability of the circular orbit. From the discussion of the previous section, it should
be clear that this is a necessary but not sufficient condition for the space-time to be a BST. Specialising to the case where
$g_{tt} = -e^{2\phi(q)}$, $g_{qq} = 1/\beta^2$, we find that the conserved energy and angular momentum for circular orbits can be determined as
\begin{equation}
E^2 = \frac{e^{2\phi(q)}}{1-q\phi'(q)}~~,~~L^2 = \frac{q^3 \phi'(q)}{1-q\phi'(q)}
\end{equation}
Positivity of these quantities demand that 
\begin{equation}
0<q\phi'(q)<1
\label{encon}
\end{equation}
for all values of $q$, which is the same result derived in \cite{Perlick} in a 
slightly different way. Further, in order to satisfy $V''(q) >0 $, we require at all points,
\begin{equation}
\frac{2 \beta ^2 \left(-r \phi ''(q)+2 r \phi '(q)^2-3 \phi '(q)\right)}{q \left(q \phi '(q)-1\right)} > 0
\end{equation}
As mentioned earlier, this has to be complemented by the validity of the weak energy conditions, and the fact that our solution should
be matched to an external Schwarzschild spacetime. This latter fact means that the radial pressure should be zero for some real value of
the radial coordinate. 

Simultaneous analysis of the conditions mentioned above with a general form for $\phi(q)$ is 
rather cumbersome and not particularly illuminating. To glean meaningful insight, we will consider some simple logarithmic 
forms of $\phi(q)$. We first consider 
$e^{2\phi (q)}=\frac{1}{Aq^m+Bq^n}$. Here, we assume that $A$ and $B$ are 
non-negative constants since this function should be positive for all values of $q$. At the boundary ($q=q_b$), where the metric is 
matched to a Schwarzschild space-time, we have 
$e^{2\phi(q_{b})}=\frac{1}{Aq_{b}^m+Bq_{b}^n}=\beta^2$. 
We can calculate the $P_{r}$ in this case, and it turns out to be
\begin{equation}
P_{r}=-\frac{1}{q^2}\left[\frac{A\lbrace(m-1)\beta^2+1\rbrace q^{m-1}+B\lbrace(n-1)\beta^2+1\rbrace q^{n-1}}{Aq^{m-1}+Bq^{n-1}}\right]~.
\label{genpr}
\end{equation}
If $P_{r}=0$ always, then eq.(\ref{genpr}) is seen to reduce to $m=n=\left(1-\frac{1}{\beta^2}\right)=\frac{\beta^2-1}{\beta^2}$. 
The constant $A$ can then be evaluated from the matching condition at the boundary, and finally the space-time metric can be written as :
\begin{equation}
ds^2=-\beta^2\left(\frac{q}{q_{b}}\right)^{\frac{1-\beta^2}{\beta^2}}dt^2+\frac{dq^2}{\beta^2}+q^2d\Omega^2~,
\label{jmnst}
\end{equation}
which is the JMN space-time \cite{JMN1}, that has a naked singularity at the center and the radial pressure is zero. Note that $\rho \sim 1/q^2$ and
diverges at the origin. 

In this case, the weak energy conditions are always satisfied, and since the radial pressure is zero, the space time can be matched with
an external Schwarzschild solution for all values of the radial coordinate. However, we note that the condition for existence of stable circular orbits 
at all points given in  eq.(\ref{encon}) translates (with $0<\beta<1$) into $\beta > 1/\sqrt{3}$, which is a stronger constraint than 
$\beta > 1/\sqrt{5}$ derived in \cite{JMN1}.

Alternatively, let us take $m=0$ and $n= -1$ then the space-time metric is 
\begin{equation}
ds^2=-\frac{1}{A+\frac{B}{q}} dt^2+\frac{dq^2}{\beta^2}+q^2d\Omega^2~,
\end{equation}
This is an example of a Bertrand spacetime studied by Perlick in \cite{Perlick} and as mentioned before, one can show that here the circular orbits
are closed under small perturbations. In \cite{BST1}, \cite{BST2}, it has been shown that BSTs, thought of as galacic metrics give rise
to excellent fits to data on galactic rotation curves for low surface brightness galaxies. For this spacetime, we get the radial pressure as
\begin{equation}
P_{q}=\frac{\beta^2(2B+Aq)-(Aq+B)}{q^2(Aq+B)}~.
\end{equation}
The above space-time can be written as:
\begin{equation}
ds^2=-\frac{2\beta^2}{1+\frac{q_{b}}{q}}dt^2+\frac{dq^2}{\beta^2}+q^2d\Omega^2~,
\label{BST}
\end{equation}
where the space-time has been matched with Schwarzschild space-time at $q=q_{b}$. For the metric defined by eq.(\ref{BST}), $\rho \sim 1/q^2$ and 
this is again indicative of a naked singularity at the center. 

In this case, we find that the weak energy condition is always satisfied, and so is the condition for the existence of stable circular orbits at
all points. However, the condition that the radial pressure is zero at a finite positive value of the radial coordinate shows that 
$\beta > 1/\sqrt{2}$. 

It should be noted that although both the metrics of eq.(\ref{jmnst}) and eq.(\ref{BST}) have closed circular orbits at all points, 
only that of eq.(\ref{BST}) falls under the category of BSTs, i.e the circular orbits are {\it closed} under small 
perturbations. This can be checked by comparison with the generic BST metrics derived in \cite{Perlick}.\footnote{Various properties of 
BSTs, including aspects of galactic rotation curves and gravitational lensing phenomena have been studied 
in \cite{BST1}, \cite{BST2}, where it was shown that they can be thought of as realistic 
galactic models, upon comparison with existing experimental data.} Importantly, our analysis shows that it is possible to obtain
space times, where closed, stable orbits can exist at all values of the radial coordinates, via a collapse process.  This is the main 
result of this subsection. 

We should also point out that apart from the cases considered here, there are several other possible metrics that satisfy the 
criteria specified above. In particular, we could have various forms of $e^{2\phi(q)}$ of eq.(\ref{finalmetric}) other than the ones
considered here. In particular, one could also have general forms such as $e^{2\phi(q)} = \sum_i A_i q^{p_i}$ of $1/ \sum_i A_i q^{p_i}$ where
$A_i$ are arbitrary coefficients. Considering equation (\ref{encon}) in the large and small $q$ limits, it can be checked that one should 
have $|p_i| \leq 2$. Here, we have simply discussed the simplest possibilities, and the physical relevance or motivation for more general 
solutions are not clear to us, and will not be discussed further.  

As mentioned in the introduction, our purpose now would be to understand the nature of fluids that source the space-times discussed
in this subsection. This is the analysis that we currently undertake. 

\section{Relativistic two-component fluid models for naked singularities}

In the discussion of the previous subsection, we have considered collapse scenarios that incorporate anisotropy, i.e the
radial and tangential pressures are unequal. If we think of the possible nature of fluids that source such space times, these
will thus be anisotropic. 
Importantly, as we have extensively mentioned, our aim is to model galactic dark matter arising out of a collapse process. Hence 
the analysis below pertains to two-fluid models of anisotropic galactic dark matter. 

In general relativity, one can model an anisotropic fluid, for which the principle pressure components are not
identical, as a composition of two perfect fluids. The formalism for this analysis has been developed in \cite{Letelier}, \cite{Bayin}
and we will closely follow the notations used in these works. To match with standard notation, we will also call the radial coordinate
as $r$ (which is the same as $q$ in the previous subsection). 

For a generic space-time with metric tensor $g_{\mu\nu}$, the total energy-momentum
tensor of the two non-interacting perfect fluids can be written as
\begin{equation}
T_{\mu\nu}=(\rho_1+P_1)u_{\mu}u_{\nu}+g_{\mu\nu}P_1+(\rho_2+P_2)v_{\mu}v_{\nu}+
g_{\mu\nu}P_2
\end{equation}
where $u_{\mu},v_{\mu}$ are the 4-velocities of the two component fluids such that $u_{\mu}u^{\mu}=v_{\mu}v^{\mu}=-1$.
Here, $\rho_1,P_1$ and $\rho_2,P_2$ are their densities and pressures of the two fluids, respectively. 

Now, after a linear transformation in the 4-velocity space, it can be shown that the energy-momentum tensor can be
can be expressed as \cite{Letelier}
\begin{eqnarray}
T_{\mu \nu} = (\rho + p_\perp)w_\mu w_\nu + p_\perp g_{\mu \nu} +
(p_r - p_\perp) y_\mu y_\nu\,,
\label{2frmt1}
\end{eqnarray}
where 
$$w^\mu w_\mu =-1\,,\,\,\,y^\mu y_\mu=1\,,\,\,\,w^\mu y_\mu =0\,,$$ where $w_\mu$ is the 4-velocity of the effective two-fluid
system and $y_\mu$ denotes a spacelike vector along the anisotropy direction. The energy density $\rho$ and pressure $p$ appearing in
eq.(\ref{2frmt1}) can be shown to be given, in terms of $\rho_1,P_1$ and $\rho_2,P_2$ as \cite{Letelier},\cite{Bayin}
\begin{eqnarray}
\rho &=& -\frac12(\rho_1 - P_1 + \rho_2 - P_2) + \frac12 \left[
(P_1 + \rho_1 + P_2 + \rho_2)^2\right.\nonumber\\ 
&+&\left. 4(P_1 + \rho_1)(P_2 + \rho_2)
\left\{(u_\mu v^\mu)^2 -1\right\}\right]^{1/2}\,,
\label{rhot}\\
P_r &=&\frac12(\rho_1 - P_1 + \rho_2 - P_2) + \frac12 \left[
(P_1 + \rho_1 - P_2 - \rho_2)^2\right.\nonumber\\ 
&+&\left. 4(u_\mu v^\mu)^2(P_1 + \rho_1)(P_2 + \rho_2)\right]^{1/2}\,,
\label{prt}\\
p_\perp &=& P_1 + P_2\,.
\label{pperpt}
\end{eqnarray}
Now we require to make a specific choice of coordinates, and  for spherically symmetric anisotropic fluids, we choose
\begin{eqnarray}
y^0=y^2=y^3=0;~~~~~w^1=w^2=w^3=0
\end{eqnarray}
such that $y^1y_1= 1$ and $w^0w_0=-1$. Then one can write the energy momentum tensor of eq.(\ref{2frmt1}) as 
\begin{equation}
T^0_0 = -\rho,~~~T^1_1 = p_r,~~~T^2_2 = T^3_3 = p_{\perp}
\end{equation}
One can then use the Einstein's equation, and denoting the Einstein tensor by $G_{\nu}^{\mu}$, we have
\begin{eqnarray}
-G_0^0 = \rho,~~~ G_1^1 = p_r, ~~~ G_2^2 = p_1 + p_2
\label{main2fluid}
\end{eqnarray}
where $G_0^0$, $G_1^1$ and $G_2^2$ can be obtained from the metric of eq.(\ref{finalmetric}), respectively, and
$\rho$, $p_r$ and $p_{\perp}$ are given by eqs.(\ref{rhot}), (\ref{prt}) and (\ref{pperpt}) respectively.  Denoting $u_\mu v^\mu =
K$, it is then seen that we have three equations for five unknowns, i.e $$\rho_1(r)\,,\,\rho_2(r)\,,\,P_1(r)\,,\,P_2(r)\,,\,K(r)$$ defining
the two-fluid system These will yield a solution if we fix barotropic equations of state for
the component fluids, and we assume
\begin{eqnarray}
P_1 = K_1 \rho_1\,,\,\,\,\,\,P_2=K_2 \rho_2
\label{eos2}
\end{eqnarray}
where $K_1$ and $K_2$ are real constants. Here, we will restrict the values of the constants to be 
\begin{eqnarray}
-1 < K_1\,,\,K_2 < 1
\label{gamval}
\end{eqnarray}
This can be interpreted as a restriction on the nature of the component fluids. The lower limit of
the $K$'s specify that we are neglecting phantom fields from the matter part and the upper limit indicates that the equations 
of state does not become too steep.  

Now we can use eq.(\ref{eos2}) and eqs.(\ref{rhot}) - (\ref{pperpt}),
and after some algebra, we get the following expressions (these can be easily checked by
explicit substitution from eq.(\ref{main2fluid}), and eqs.(\ref{rhot}) - (\ref{pperpt})).
\begin{eqnarray}
\rho_1(r) &=& \frac{K_2\left(G_2^2 - G_1^1 - G_0^0\right)-G_2^2}
{(K_2 - K_1)}\,,
\label{r1f}\\
\rho_2(r) &=& \frac{K_1(G_0^0+G_1^1-G_2^2) + G_2^2}{
(K_2 -K_1)}\,,
\label{r2f}\\
u_{\mu}v^{\mu} \equiv K &=& 
-\left[\frac{\left[G_1^1 + (\rho_1 - K_2\rho_2)\right]
\left[G_1^1 + (\rho_2 - K_1\rho_1)\right]}
{ \rho_1 \rho_2 (1+K_1)(1+K_2)}\right]^{1/2}
\label{umuvmu}
\end{eqnarray}
The negative sign in the last equation is due to the fact that $u$ and $v$ are time-like vectors.  

We now apply this formalism to the JMN space-time of eq.(\ref{jmnst}). In that case, we find
\begin{eqnarray}
\rho_1 &=& \frac{\left(\beta ^2-1\right) \left(\beta ^2 \left(3
   K_2+1\right)+K_2-1\right)}{4 \beta ^2 \left(K_1-K_2\right) r^2} \nonumber\\
\rho_2 &=& -\frac{\left(\beta ^2-1\right) \left(\beta ^2 \left(3
   K_1+1\right)+K_1-1\right)}{4 \beta ^2 \left(K_1-K_2\right) r^2}
\label{K1K2JMN}
\end{eqnarray}
The expression for $K$ is somewhat lengthy and we omit it here. From eq.(\ref{K1K2JMN}), it can be seen that only in the limit
$\beta \to 1$, positivity of the energy densities require that $K_1$ and $K_2$ be of opposite signs, i.e
$K_1K_2 <0$ (negativity of $K$ also  yields the same result in this limit). However, this limit is problematic, as from eq.(\ref{jmnst}) it
can be seen that the space-time then reduces to flat space, with all the component of the energy-momentum tensor vanishing, i.e
no fluid description is possible. This limit therefore has to be ruled out. Since no such statement can be made for other values of 
$\beta$ (remembering that $\beta > 1/\sqrt{3}$, following our discussion in the previous subsection), we can conclude that for 
JMN space-times, $K_1$ and $K_2$ can be of same (or opposite) signs. 

Next, let us consider the metric of eq.(\ref{BST}). This case has been studied in \cite{BST1} and we will be brief here. We simply
record the expression for the ratio of the energy density of the component fluids, given by\footnote{Both $\rho_1$ and $\rho_2$ $\sim
\frac{1}{r^2}$ for large $r$ in BSTs \cite{BST1}.}
\begin{equation}
\frac{\rho_1}{\rho_2} = \frac{\beta ^2 \left(K_2 \left(8 r^2+22 r+11\right)-2 r+1\right)-8 K_2
   (r+1)^2}{8 K_1 (r+1)^2-\beta ^2 \left(K_1 \left(8 r^2+22 r+11\right)-2
   r+1\right)}
\end{equation}
For large $r$, it is seen that $\rho_1/\rho_2 \to -K_2/K_1$ and we see that the condition of positivity of the energy densities
necessarily require $K_1K_2 < 0$ here, irrespective of the value of $\beta$. 

\section{Non-relativistic limit of relativistic two-component fluids}

We will now set up an analysis of the non-relativistic limit of the two-component dark matter fluids discussed in the previous subsection. In order
to do this, we will need an expression for the non-relativistic limit of the density profiles for the metrics discussed above. This can 
be done if we compute the circular velocity arising out of these metrics and then interpret them as arising out of a mass distribution
in the Newtonian limit. In GR, this circular velocity is a frame dependent quantity, and can be defined only in a locally flat (tetrad) basis. 
In such a basis, one has to project the four momentum of the particle onto the tetrad frame time axis, and then
equate this to the Lorentzian expression for the energy \cite{Hartle}. This will necessitate that the locally flat observer is at the same radial 
distance as the body undergoing geodesic motion. Such a definition has been used in the GR literature, but it is difficult to reconcile this 
with results from galactic rotation curves, which are usually obtained as a function of a radial distance, since from the GR point of 
view this would require a series of Lorentzian observers at different radii.

On the other hand, in \cite{BST1}, we had proposed a purely phenomenological definition of the circular velocity that can be computed
via GR as 
\begin{equation}
v_{circ} = r\frac{d\phi}{dt} \equiv r\frac{L}{E}\frac{g_{tt}}{g_{\phi\phi}}
\label{vcirc}
\end{equation}
with the second identity following from the definition of $E$ and $L$ given earlier (see discussion following eq.(\ref{genmotion})). 
This definition of the circular velocity was used in \cite{Roberts}.
In \cite{BST1}, this definition of $v_{circ}$ was shown to match well with experimental data for various low surface brightness galaxies.
With the (phenomenological) definition of the circular velocity in eq.(\ref{vcirc}), we can compute the Newtonian mass density as
\begin{equation}
\rho_N(r) \sim \frac{1}{r^2}\frac{d}{dr}\left(v_{circ}^2 r\right)
\end{equation}
where we have ignored factors of $\pi$ on the right hand side, and the subscript $N$ refers to the Newtonian limit being considered here.

Let us first consider the JMN space-time given in eq.(\ref{jmnst}). In this case, a simple computation yields 
\begin{equation}
v_{circ}^{JMN} = \sqrt{\frac{1-\beta^2}{2}}r^{\frac{1-\beta^2}{2\beta^2}} \implies \rho_N^{JMN} \sim \frac{1}{r^{\left(3-\frac{1}{\beta^2}\right)}}
\label{JMNden}
\end{equation}
On the other hand, let us consider the BST metric of eq.(\ref{BST}). In this case, we find that 
\begin{equation}
v_{circ}^{BST} = \frac{\beta\sqrt{r}}{1+r} \implies \rho_N^{BST} \sim \frac{\beta^2}{r\left(1+r\right)^3}
\label{BSTden}
\end{equation}

From eq.(\ref{JMNden}), we see that the falloff of the Newtonian mass density is $\sim r^{-n}$, with $n < 2$ (since
$0<\beta<1$). On the other hand, for BSTs, the large distance falloff is $\sim r^{-4}$. Recalling our analysis of the equations
of state for relativistic two-component fluids of the last subsection, this is {\it indicative} of the fact that when the falloff of the mass 
density is with a (negative) power that is less than $2$, the
two component fluids can have barotropic equations of state ($P_1 = \rho_1 K_1, P_2 = \rho_2 K_2$) where $K_1$ and $K_2$ can
be of the same (or opposite) signs. Although we have only analysed an $\sim r^{-4}$ fall-off of the mass density, 
it naively seems that the previous statement is no longer true when the falloff is with a (negative) power 
greater than $2$ where we necessarily have $K_1K_2<0$. The case $n = 2$ (i.e $\beta \to 1$ in eq.(\ref{JMNden})) is problematic, as this
necessitates that the circular velocity is zero, as are all component of the energy momentum tensor (recall the discussion towards the
end of the last subsection). 

Here, we have taken the non-relativistic limit of a GR result. In the next section, we complement this by analysing a purely Newtonian 
two-component fluid and compare it with this analysis.  

\section{Newtonian two-component fluids}

In this section, we consider a purely Newtonian two-fluid model 
of galactic dark matter, of two different fluids in equilibrium, interacting only
via Newtonian gravity and differing in their equations of state. The motivation for this analysis comes from the fact that
in the last section, we considered the non relativistic limit of the two-fluid models that describe the space-times of
eqs.(\ref{jmnst}) and (\ref{BST}), and saw that there were some non-trivial constraints on the equation of state parameters. 
Here, we will analyse if such constraints appear in the purely Newtonian case, and this should complement the results
of the previous section. 

We should point out the important assumptions that we make in this section. 
\begin{itemize}
\item
The fluids are assumed to be in steady state, i.e their properties are independent of time. 
\item
We treat the fluids phenomenologically, i.e we {\it assign} simple velocities to the component fluids and analyse the possible constraints on
the resulting equation of state (we momentarily elaborate on this). 
\item
We will ignore the effects of energy dissipation in our analysis, i.e effects of viscosity etc. are assumed to be small.
\item
A spherically symmetric situation has been assumed, i.e all the variables that enter our computations are functions of only the radial coordinate.
\item
Thermal equilibrium of baryonic matter with the fluids that we have dealt with here are not considered.
\end{itemize}

Of course, each of these assumptions can be questioned on grounds of physicality in dealing with realistic systems, but we will proceed 
with the understanding that these simplify the analysis while allowing an analytical handle on the physics of the system. Indeed, it will be 
interesting to relax one or more of these assumptions, in which case a more sophisticated numerical analysis than the ones performed in 
this paper needs to be invoked.

As mentioned, our analysis here will be entirely phenomenological : instead of a first principles computation of the fluid properties with appropriate
boundary conditions, we will here assign different velocities to the two fluids, and then try to understand the possible constraints
on the equations of state that might arise. As we show below, this will enable us to retain an analytic handle on the models, while allowing a
comparison with the relativistic analysis of the previous section. We do not claim any generality of these results beyond the simple 
situations considered here. 

The two fluids that we consider are self-gravitating, and also
mutually interacting via gravity, although they independently conserve momenta by following two independent
Navier-Stokes equations.  The system of equations governing two independent fluids in their gravitational fields are given by two
continuity equations, a set of six Navier-Stokes equations, the Poisson equation for gravity and two independent equations of state 
which we choose to be polytropic (in particular barotropic, to compare with the results of the previous section).  

The continuity equations for the two fluids become
\begin{eqnarray}
\frac{\partial \rho_{i}}{\partial t} + \nabla \cdot (\rho_{i}{\bf v}_{i})=0\,,
\label{conte}
\end{eqnarray}
where $i=1,2$ denote the two component fluids.\footnote{This will be our notation convention 
throughout this section and we will not mention this further.} The Navier-Stokes equations now become 
\begin{eqnarray}
\frac{\partial {\bf v}_{i}}{\partial t} + ({\bf v}_{i}\cdot \nabla){\bf v}_{i}
= -\frac{\nabla P_{i}}{\rho_{i}} + \nu_{i}\nabla^2 {\bf v}_{i} - \nabla \Phi
\label{ns1}
\end{eqnarray}
Since we assume the two fluids to be in steady state, the time derivatives in eqs.(\ref{conte}) and (\ref{ns1}) are taken to 
be zero. Also, we assume that dissipation effects are negligible, i.e $\nu_i = 0$ ($i=1,2$) in eq.(\ref{ns1}). 
The Poisson equation for gravity may be written as
\begin{equation}
\frac{1}{r^2}\frac{d}{dr}\left(r^2\frac{d\Phi}{dr}\right)=4\pi G \rho~,~~\rho=\rho_1 + \rho_2\,.
\label{2pe}
\end{equation}
Here we assume the two fluid species to follow two independent polytropic equations of state
\begin{eqnarray}
P_1= K_1\rho_1^{\alpha_1}~,~~P_2= K_2\rho_2^{\alpha_2}\,
\label{2poly}
\end{eqnarray}
In the non-relativistic case
the net pressure of the two fluids is simply
\begin{eqnarray}
P=P_1+P_2\,
\label{npress}
\end{eqnarray}
To connect to the results of the previous section, we will henceforth set $\alpha_1=\alpha_2=1$, i.e assume barotropic
equations of state for the component fluids. Then, one can use
the second of eq.(\ref{2pe}), along with eqs.(\ref{2poly}) and (\ref{npress}) to obtain 
\begin{eqnarray} 
\rho_1=\frac{P-K_2 \rho}{K_1 - K_2}\,,\,\,\,\,
\rho_2=\frac{P-K_1 \rho}{K_2 - K_1}\,\,\,\,\,
\label{rho12}
\end{eqnarray}
which shows that the above set of relations can be used as
constraints in such a way that $\rho_i > 0$ is always satisfied in our region of physical interest. If without loss of generality,
we consider $K_1>K_2$ then it is seen from eq.(\ref{rho12}) that 
\begin{equation}
K_1>\frac{P}{\rho}>K_2
\label{barocon}
\end{equation}
is needed to have $\rho_1,~\rho_2 >0$. If we consider the total density of the composite fluid as \footnote{Since we are primarily
interested in modelling galactic dark matter, the density profile of the composite fluid should be a generalised Hernquist 
profile \cite{Hernquist},\cite{Zhao} which justifies the form of eq.(\ref{density}).}
\begin{equation}
\rho = \frac{\rho_0}{\left(\frac{r}{R_s}\right)^{\alpha}\left(1+\frac{r}{R_s}\right)^{\beta}}
\label{density}
\end{equation}
then we have to check for what values of $\alpha$ and $\beta$ the condition in eq.(\ref{barocon}) is satisfied.
It can be checked by an asymptotic analysis in both the small and the large $r$
limits, that the condition mentioned in eq.(\ref{rho12}) is indeed satisfied in
the case of a static fluid system having an overall density profile of
the NFW ($\alpha=1$, $\beta=2$) or Hernquist ($\alpha=1$, $\beta=3$)
form and with its individual components satisfying barotropic
equations of state. If we introduce radial velocities to the component
fluids which still follow barotropic equations of state, it can be
numerically verified that the above condition necessary for
well-behaved densities is still satisfied in case the total density
profile is an NFW or a Hernquist profile.

\subsection{Two-component static fluid}

As mentioned in the beginning of this section, we will follow a phenomenological approach, and assign various
velocity profiles to our model to see the constraints that might result. 
First we consider the case $v_{ir}=v_{i\phi}=v_{i\theta}=0$ ($i=1,2$) i.e the static fluid. 
 One can derive from the  Poisson and Navier-stokes equations for the composite fluid, 
\begin{equation}
\frac{1}{r^2}\frac{d}{dr}\left(r^2\frac{d\Phi}{dr}\right)=4\pi G\rho~,~\frac{dP}{dr}=-\rho\frac{d\Phi}{dr}~.
\label{nav2}
\end{equation}
The Navier-Stokes equations for the component fluids boil down to
\begin{equation}
\frac{dP_1}{dr}+\rho_1\frac{d\Phi}{dr}=0~,~~
\frac{dP_2}{dr}+\rho_2\frac{d\Phi}{dr}=0~.
\label{3nav}
\end{equation}
Then from eq.(\ref{rho12}), using eqs.(\ref{2pe}) and (\ref{npress}), we have the following differential equations in $\rho_1$ and $\rho_2$ :
\begin{equation}
\frac{d\rho_1}{dr}=\frac{1}{K_2-K_1}\left(\rho\frac{d\Phi}{dr}+K_{2}\frac{d\rho}{dr}\right)~,
~~\frac{d\rho_2}{dr}=\frac{1}{K_1-K_2}\left(\rho\frac{d\Phi}{dr}+K_{1}\frac{d\rho}{dr}\right)~.
\label{drho}
\end{equation}
\\
Assuming that the total mass density in the asymptotic (large $r$) limit falls off as a power law, we impose 
\begin{equation}
\rho\sim\frac{1}{r^n}~,~n \geq 1~,
\end{equation}
then modulo some irrelevant constants, we have
\begin{eqnarray}
\frac{d\rho_1}{dr} = \frac{1}{K_2-K_1}\left(\frac{1}{(3-n)r^{2n-1}} - \frac{nK_2}{r^{n+1}}\right),~
\frac{d\rho_2}{dr} = \frac{1}{K_1-K_2}\left(\frac{1}{(3-n)r^{2n-1}} - \frac{nK_1}{r^{n+1}}\right)
\label{dender}
\end{eqnarray}
Let us, without loss of generality, assume that $K_2 > K_1$. Then, for $n > 2,~n\neq 3$ (we will momentarily come to the $n=3$ case), 
we see that the second terms in the parentheses of both the equations of eq.(\ref{dender}) dominate at large $r$, so that in order
to have $d\rho_{1,2}/dr <0$, we must necessarily have $K_2 > 0$ and $K_1 < 0$. 
So the condition for positive component densities is $K_1K_2<0$, which implies that one of the component fluids 
has negative pressure, i.e has properties similar to dark energy. 
The case $n=3$ is qualitatively similar to the above analysis. Here, one only has to remember that the first terms in the parentheses
of the right hand side of both the equations in eq.(\ref{dender}) are replaced by a term $\sim \log(r)/r^5$ while the second term 
for both $\sim 1/r^4$. Hence it is the second term that again dominates the right hand side of both the equations, and we again have
$K_1K_2 <0$ for physically relevant density profiles of the individual fluids. 

The $n=2$ case is qualitatively different. In this case, it is seen that 
\begin{equation}
\frac{d\rho_1}{dr} = \frac{1}{K_2-K_1}\left(\frac{1-2K_2}{r^{3}}\right),~
\frac{d\rho_2}{dr} = \frac{1}{K_1-K_2}\left(\frac{1-2K_1}{r^{3}}\right)
\label{dender1}
\end{equation}
It is easily seen that negativity of the right hand side of the two equations in eq.(\ref{dender1}) does not necessarily imply 
that $K_1K_2 < 0$, i.e we can have a solution with both fluids having positive pressure. Finally, we note that the $n<2$ case is ruled out, 
since from eq.(\ref{dender}) it is seen that in this case $d\rho_1/dr$ and $d\rho_2/dr$ can never be made simultaneously negative. 

\subsection{Two-component fluid with radial velocity}

Now we consider the case where $v_{ir}\neq 0$ and $v_{i\phi}=v_{i\theta}=0$, where $i=1,2$ denotes the individual fluids.
From the continuity equations for the two component fluids we have here
\begin{equation}
v_{ir}=\frac{A_i}{r^2\rho_i}
\label{vir}
\end{equation}
From the Poisson equation for gravity one can get the functional forms for $\Phi$ and $\frac{d\Phi}{dr}$. 
Now the radial Navier-Stokes equations for these 
two fluids boil down to 
\begin{equation}
\rho_1v_{1r}\frac{dv_{1r}}{dr}=-\frac{dP_1}{dr}-\rho_1\frac{d\Phi}{dr}~,~~
\rho_2v_{2r}\frac{dv_{2r}}{dr}=-\frac{dP_2}{dr}-\rho_2\frac{d\Phi}{dr}
\label{2nav}
\end{equation}
From eqs.(\ref{2nav}), (\ref{vir}) and (\ref{2poly}), we can get the analytical expressions, $\frac{d\rho_{1}}{dr} = \frac{{\mathcal A}_1}{{\mathcal B}_1}$, 
$\frac{d\rho_{2}}{dr}=\frac{{\mathcal A}_2}{{\mathcal B}_2}$, where 
\begin{eqnarray}
{\mathcal A}_1&=&\rho\frac{d\Phi}{dr}+\left(K_{2}-\frac{{A_2}^2}{r^4\rho_2^2}\right)
\frac{d\rho}{dr}-\frac{2}{r^5}\left(\frac{{A_1}^2}{\rho_1}+\frac{{A_2}^2}{\rho_2}\right)\nonumber\\
{\mathcal B}_1&=&\frac{1}{r^4}\left(\frac{{A_1}^2}{{\rho_1}^2}-\frac{{A_2}^2}{\rho_2^2}\right)+K_2-K_1 \nonumber\\
{\mathcal A}_2&=&\rho\frac{d\Phi}{dr}+\left(K_{1}-\frac{{A_1}^2}{r^4\rho_1^2}\right)
\frac{d\rho}{dr}-\frac{2}{r^5}\left(\frac{{A_2}^2}{\rho_2}+\frac{{A_1}^2}{\rho_1}\right)\nonumber\\
{\mathcal B}_2&=&\frac{1}{r^4}\left(\frac{{A_2}^2}{{\rho_2}^2}-\frac{{A_1}^2}{\rho_1^2}\right)+K_1-K_2
\label{d2rho}
\end{eqnarray}
We will now analyse the possible constraints on the equation of state, from eqs.(\ref{d2rho}). 

First consider the case when the falloff of the densities of the component fluids are similar, i.e we have 
\begin{equation}
\rho=\frac{\rho_0}{r^n},~\rho_1=\frac{\rho_{10}}{r^n},~\rho_2=\frac{\rho_{20}}{r^n}
\label{liminf}
\end{equation}
Here, using the asymptotic forms of 
$\rho$ and $\rho_{1,2}$ from eq.(\ref{liminf}) in eqs.(\ref{d2rho}), we have in the limit $r \rightarrow \infty$ by equating the coefficients 
of the dominant terms, 
\begin{equation}
n\rho_0\left(\frac{{A_2}^2}{{\rho_{20}}^2}-\frac{{A_1}^2}{{\rho_{10}}^2}\right)=
\frac{n{A_2}^2\rho_0}{{\rho_{20}}^2}-2\left(\frac{{A_1}^2}{\rho_{10}}+\frac{{A_2}^2}{\rho_{20}}\right)
\end{equation}
Using the expression $\rho_0=\rho_{10}+\rho_{20}$, we then have the constraint
\begin{equation}
\left(n-2\right)\left(\frac{{A_1}^2}{\rho_{10}}+\frac{{A_2}^2}{\rho_{20}}\right)=0
\end{equation}
This implies that $A_1=A_2=0$ for $n \neq 2$. From eq.(\ref{vir}), this will correspond to the case where both the fluids
have zero velocity, i.e are static. Hence the results of the previous subsection can be applied in this case, and we would necessarily
have $K_1K_2<0$, akin to that example, and from that discussion, it also follows that the case $n < 2$ is ruled out on physical grounds. 

The case $n=2$ is qualitatively different. We shall not belabour the details here, but simply state the result that a careful analysis reveals that
that the product $K_1K_2$ need not always be negative, unlike the case $n > 2$ (the case $n < 2$ is anyway ruled out). This is analogous to 
the static fluid case considered in the previous subsection. 

Hence we can conclude that if the large $r$ behaviour of the densities of the two fluids is similar, then we can have non-zero radial velocities 
in the system only if the total density profile is that of an isothermal sphere. In that case however, we are not guaranteed to have $K_1K_2<0$. This
last condition holds for all other profiles, with $n>2$, in which case we are constrained to have a static two-fluid solution, as explained, where the
velocity components of both fluids vanish identically. 

Finally, we have analysed the case where the asymptotic falloff of the two component fluids might be according to different power laws, i.e we
assume that at large values of the radial coordinate, $\rho_1 \sim \frac{1}{r^n}$ and $\rho_2 \sim \frac{1}{r^m}$ so that assuming without 
loss of generality $m > n$, we have the density of the composite fluid $\rho \sim \frac{1}{r^n}$. The analysis is somewhat cumbersome and
we will only state the main results here. After a detailed analysis, we find from eq.(\ref{d2rho}) that consistency of the falloff behaviour 
of the component fluids necessarily means that the component fluid with the steeper falloff must have zero radial velocity and that 
for $n>2,~m>n$, we again get back the condition $K_1K_2<0$. For $n=2,m>n$, this condition need not be strictly satisfied. 

Now we will compare this with the results of the Newtonian limit of the relativistic fluid considered in subsection 2.3. There, we found 
{\it indication} that if the fall-off of the mass density is $\sim r^{-n}$ with $n>2$, it necessary implies that either component of the
composite fluid should have exotic behaviour, which was not the case for $n<2$. The $n=2$ case was ruled out due to the nature of the
JMN space-time. In the Newtonian analysis, we find that for the simple fluid models that we have considered, if again the fall-off of the
mass density is asymptotically  $\sim r^{-n}$ with $n>2$, then this necessarily leads to negative pressure for one of the component fluids. 
The $n<2$ region could not be studied due to the limitations of our simplistic models, and the case $n=2$ showed that none of the
fluids need to be exotic in this case. By comparison, a definitive conclusion regarding fluid behaviour can be reached only for the case
$n>2$, in which case we may conclude that one of the fluids must have negative pressure.

\section{Conclusions and future directions}

In this paper, we first studied gravitational collapse in an anisotropic scenario. We have shown that there is a class of space-times 
that can arise in this case, of which the JMN and the BST are special cases. We have further showed that if the BST is thought of as
a two-fluid model, then one of the fluids must have negative pressure. A similar analysis was then performed for Newtonian two component
fluids, with some important simplifying assumptions. 

In the relativistic setup, we saw that anisotropy dictates that we use a two-fluid model to describe the energy momentum tensor of 
the space-time formed after gravitational collapse. In that case, we used barotropic equations of state, $P_1 = \rho_1K_1$
and $P_2 = \rho_2K_2$ for the individual fluids and saw
that for BSTs, the pressure of one of the fluids must necessarily be negative. This was not the case with the JMN space-time. To 
put this result in perspective, we computed the mass density in a non-relativistic limit, after obtaining an expression for the circular velocity in
these space-times. The analysis was indicative of the fact that such negative pressures might arise if the falloff of the mass density is 
$\sim 1/r^n$ with $n > 2$. This is important, as it indicates that if we consider the space-time arising due to gravitational collapse as 
describing a dark matter fluid, then one of the components of the fluid might be exotic, i.e have negative pressure akin to dark energy. 

To reconcile this result, we addressed some two fluid models in a purely Newtonian setup by assuming some simple velocity profiles for
the component fluids. This analysis was purely phenomenological in nature, and we explored the constraints in the equations of state of the
component fluids in some simple two fluid models where the component fluids were {\it assumed} to have given velocities. In
particular, we first considered the case where both the fluids were static, and then the case when they could have a radial velocity. 
Our Newtonian analysis shows that in both the cases, if the fall off of the composite fluid $\sim 1/r^n$ with $n > 2$, then one of the component
fluids might have negative pressure, as was the case with the non-relativistic limit of the relativistic two fluid models. A definitive conclusion 
could not be drawn for other values of $n$.  
\begin{table}[ht]
\begin{center}
 \begin{tabular}{|p{4.5cm}|p{4cm}|p{4.5cm}|}
\multicolumn{3}{c}
{Table 1} \vspace{0.3cm}\\
\hline
Two fluid system & Density falloffs  &The condition $K_1K_2 <0$  \\
\hline
JMN (relativistic) & $\rho_1,\rho_2 \sim 1/r^2$ & Not always satisfied \\
\hline
BST (relativistic) & $\rho_1,\rho_2 \sim 1/r^2$ &Always satisfied \\
\hline
JMN (non-relativistic) & $\rho \sim 1/r^n$, $n<2$ &Not satisfied (Indicative) \\
\hline
BST (non-relativistic) &$\rho \sim 1/r^4$ & Satisfied (Indicative)\\
\hline
Newtonian fluid &$\rho_1,\rho_2 \sim 1/r^n$, $n>2$ &Always satisfied \\
\hline
Newtonian fluid &$\rho_1,\rho_2 \sim 1/r^n$, $n=2$ & Not always satisfied \\
\hline
Newtonian fluid &$\rho_1 \sim 1/r^n,~\rho_2 \sim 1/r^m$, $n>2,~m>n$ &Always satisfied \\
\hline
Newtonian fluid &$\rho_1 \sim 1/r^n,~\rho_2 \sim 1/r^m$, $n=2,~m>n$ &Not always satisfied \\
\hline
\end{tabular}
\caption{Validity of the constraint $K_1K_2 < 0$ on barotropic equations of state of the form $P_1 = \rho_1K_1$, $P_2 = \rho_2K_2$, 
in two fluid models.} 
\label{table1}
\end{center}
\end{table}

We summarize this discussion in table (\ref{table1}), where the last column indicates whether the constraint $K_1K_2<0$ is 
satisfied or not. Note that the last two rows contain the results with different power law falloffs of the component fluids, that has been 
briefly discussed towards the end of subsection 5.2. 

Admittedly, there were a number of simplifying assumptions in our Newtonian analysis and the result should not be thought of as a generic
one. However, within the caveats mentioned in the paper, the results presented here seem to be indicative of some generic features of 
dark matter fluids, and this certainly deserves further study.

\end{document}